\documentclass[dvips,twocolumn,aps,showpacs,amssymb,10pt]{revtex4}
\usepackage{graphicx}
\usepackage{dcolumn}
\usepackage{bm}
\usepackage{epsfig}
\usepackage{color}
\usepackage{longtable}
\usepackage{soul}
\usepackage{ulem}
\usepackage{amssymb,amsmath}
\usepackage{verbatim}
\everymath=\expandafter{\the\everymath\displaystyle}
\newcommand{\ds}{\displaystyle}

\newcommand{\bra}[1]{\mathinner{\langle{#1}|}}
\newcommand{\ket}[1]{\mathinner{|{#1}\rangle}}

\newcommand{\redbra}[1]{\langle{#1}|\!|}
\newcommand{\redket}[1]{|\!|{#1}\rangle}

\newcommand{\dd}{\mathrm{d}}

\newcommand{\uvector}[1]{\hat{\bm{#1}}}
\renewcommand{\vec}[1]{\bm{#1}}

\newcommand{\sumint}[1]{\sum_{#1} \hspace{-0.5cm}\int\,}
\def\etal{{\it et al.}}                                        %
\begin{document}
\preprint{}
\title{Relativistic polarization analysis of Rayleigh scattering by atomic hydrogen} 
%
%

\author{L. Safari$^{1,}$\footnote{laleh.safari@oulu.fi}, P. Amaro$^{2,3}$, S. Fritzsche$^{1,4}$, J. P. Santos$^{3}$, S. Tashenov$^{2}$ and F. Fratini$^{1, 5}$} 

\affiliation{\it
$^1$ Department of Physics, University of Oulu, Box 3000, FI-90014 Oulu, Finland\\
$^2$ Physikalisches Institut, Universit\"{a}t Heidelberg, D-69120 Heidelberg, Germany\\
$^3$ Centro de F\'{i}sica At\'{o}mica, Departamento de F\'{i}sica, Faculdade de Ci\^{e}ncias e Tecnologia, FCT, Universidade Nova de Lisboa, P-2829-516 Caparica, Portugal\\
$^4$ GSI Helmholtzzentrum f\"{u}r Schwerionenforschung, D-64291 Darmstadt, Germany\\
$^5$ Departamento de F\'isica - CP 702 - Universidade Federal de Minas Gerais - 30123-970 - Belo Horizonte - MG - Brazil
} 
\date{\today \\[0.3cm]}%
%
%
%
\begin{abstract}
A relativistic analysis of the polarization properties of light elastically scattered by atomic hydrogen is performed, based on the Dirac equation and second order perturbation theory.
The relativistic atomic states used for the calculations are obtained by making use of the finite basis set method and expressed in terms of $B$ splines and $B$ polynomials.
We introduce two experimental scenarios in which the light is circularly and linearly polarized, respectively. For each of these scenarios, the polarization-dependent angular distribution and the degrees of circular and linear polarization of the scattered light are investigated as a function of scattering angle and photon energy. Analytical expressions are derived for the polarization-dependent angular distribution which can be used for scattering by both hydrogenic as well as many-electron  systems. Detailed computations are performed for Rayleigh scattering by atomic hydrogen within the incident photon energy range 0.5 to 10 keV. Particular attention is paid to the effects that arise from higher (nondipole) terms in the expansion of the electron-photon interaction.
\end{abstract}

\pacs{32.80.Wr, 32.90.+a} 
\maketitle

%
%
%
%
%
\section{Introduction}\label{sec:intoduc}

Polarization is one of the main characteristics of light which can be employed in order to investigate the properties of matter. In atomic physics, especially, the polarization properties of light have been studied for various processes, such as atomic and ionic photoionization \cite{L. Sharma:10}, hyperfine-quenched transitions \cite{A. Bondarevskaya:10}, two-photon decay \cite{Fratini:11a,F.Fratini:11_2}, atomic-field bremsstrahlung \cite{S. Tashenov:11}, radiative electron capture \cite{S. Tashenov:06} and elastic scattering of light. Elastic scattering (so called Rayleigh scattering) by atoms and ions \cite{C.F.Bohren:85,P.P.Kane:1986} has applications in astronomy, shielding, medical diagnostics and also is used extensively to obtain information about the structural properties of materials and complex molecules \cite{A. Böke:11,N.S.Kampel1:12,D.G.Norris:12,L. V. Kulik:12,V. Bellani:11,J. Crassous:12}.

Owing to the develepment of x-ray polarization sensitive detectors \cite{S. Tashenov:06, S. Tashenov:09}, tunable polarization free-electron lasers \cite{C.Spezzani:11} as well as synchrotron radiation sources \cite{F. Smend:87}, an increasing demand for accurate theoretical prediction on polarization-dependent atomic phenomena has been pointed out in the literature. For instance, elastic scattering experiments with a nearly 100\% polarized hard X-ray beam have been recently performed in ESRF (Grenoble, France), while further experiments are planned to be realized at the DESY synchrotron facility (Hamburg, Germany) in the near future \cite{Bondarev}. Moreover, the novel linear polarimetry technique based on Rayleigh scattering was used for the first time \cite{S. Tashenov:09} and recently applied as a complementary techninque to Compton scattering \cite{S. Tashenov:11}.

During the past decades, the total cross section for Rayleigh scattering has been widely investigated within the relativistic as well as nonrelativistic frameworks \cite{M.Gavrila:67, H.R.Sadeghpour:92, Nganso:07, A. Costescu:2011, A. Costescu:94, A. Costescu:2007, J.P.J.Carney:2000}. In contrast, only a few studies have been done on the angular and polarization properties of scattered photons. For example, Roy \etal{} calculated linear polarization effects in the elastic scattering of x-rays and $\gamma$-rays from targets with Z ranging from 13 to 92, by using numerically obtained Rayleigh scattering amplitudes \cite{S.C.Roy-sarkar:86}. More recently, Manakov \etal{} analyzed photon-polarization effects in two photon bound-bound atomic transitions, including relativistic and retardation effects \cite{N.L.Manakov:00}. In particular, they numerically investigated \textit{circular dichroism} effects in scattering of hard photons by hydrogenic as well as many-electron systems \cite{N.L.Manakov:1987,N.L.Manakov-Pratt:00}. However, to the best of our knowledge, a fully relativistic treatment of the polarization-dependent angular distribution in Rayleigh scattering by atomic hydrogen has not been performed yet.

In the present work, we investigate the polarization properties in Rayleigh scattering by an unpolarized hydrogen atom, within the fully relativistic framework of the Dirac equation. 
We introduce two experimental scenarios in which the light is circularly and linearly polarized, respectively. For each of these scenarios, we investigate the polarization-dependent angular distribution (PDAD) and the degrees of circular and linear polarization of the scattered light. We derive an analytical expression for the PDAD, which is valid for scattering by hydrogenic as well as many-electron systems. With the aid of the Wigner-Racah algebra, we write such expression in terms of angular parts and reduced matrix elements, where the latter are independent of the scattering geometry. The numerical evaluation of the reduced matrix elements is relativistically carried out for the Rayleigh scattering by atomic hydrogen, through the use of finite basis sets for the Dirac equation constructed from B splines and B polynomials. The photon energy range we investigate is 0.5 to 10 keV. It has been recently showed by us that, within this energy range, the finite basis set approach to Rayleigh scattering gives good agreement with NIST (National Institute of Standards and Technology) data values and other calculations \cite{L.Safari:12}.

This article is structured as follows. In Sec.~\ref{sec:Geometry}, we introduce the geometry and the notation used. In Section \ref{sec:Theory}, we recall the general polarization-dependent transition amplitude for Rayleigh scattering and evaluate it separately for the circular and linear polarization scenarios. In Section \ref{sec:Comput}, we describe the numerical method used for carrying out the calculations, while, in Sec.~\ref{sec:ResultsDiscussion}, we present the results for the PDAD and the degrees of circular and linear polarization of the light scattered by atomic hydrogen. Finally, a short summary is given in Sec.~\ref{sec:SumConcl}.

SI units are used throughout the article.

%
%
%
%
%
\section{Atomic system and geometry}\label{sec:Geometry}
\begin{figure}[b]
\includegraphics[scale=0.83]{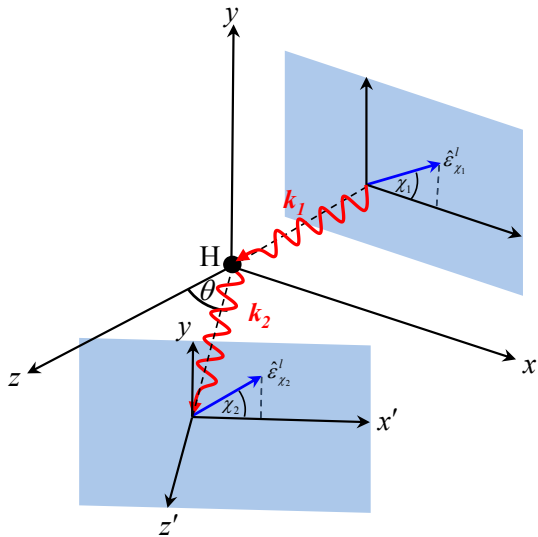}
\caption{(Color online) The adopted geometry for the scattering process is displayed. The polar angle $\theta$ uniquely defines the direction of the scattered photon in the $xz$ plane (scattering plane). The angle $\chi_1$ ($\chi_2$) parametrizes the linear polarization of the incident (scattered) photon. The hydrogen atom is placed at the origin of the coordinate axes $xyz$.}
\label{Rayleigh geometry}
\end{figure}
Let us start by introducing the atomic system and the geometry under which the distribution of the Rayleigh-scattered photons is investigated. We consider hydrogen atom in the ground state which is irradiated by light, as displayed in Fig.~\ref{Rayleigh geometry}. We adopt the quantization ($z$) axis along the direction  of the incident photon ($\vec{k}_1$). As we will see, such a choice of quantization axis simplifies the multipole expansion of the electron-photon interaction operator. The scattered photon propagates along the direction $\vec{k}_2$ at angle $\theta$ with respect to the $z$ axis. The scattering plane ($xz$) is defined by the incident and scattered photon's directions. 
Incident (scattered) photon has energy $E_{\gamma_{1 (2)}}= \hbar \omega_{1 (2)}$, propagation vector $\vec{k}_{1 (2)}$ and polarization unit vector $\uvector{\bm{\epsilon}}_{1 (2)}$, where $\hbar$ is the reduced Planck constant.

The photon polarizations we consider are linear ($\uvector\epsilon=\uvector\epsilon_\chi^l$) and circular ($\uvector\epsilon=\uvector\epsilon_\lambda^c$). Neither linear combinations of these polarizations nor mixed polarization states are taken into account. In particular, we do not consider elliptically polarized light. From an experimental point of view, linear polarization is probably the most interesting type for the considered photon energy range \cite{S. Tashenov:06}. The polarization angle $0 \le \chi \le \pi$ and the helicity $\lambda = \pm 1$ are the variables used  to parametrize the linear and circular polarization states, respectively. In this notation, $\lambda = +1$ describes right-handed and $\lambda = -1$ left-handed polarized photons, respectively. The definitions of the polarization angles are displayed also in Fig.~\ref{Rayleigh geometry} if incident and scattered photons are taken to be linearly polarized.
%
%
%
%
%
%
%
\section{Theory}
\label{sec:Theory}
\subsection{Transition amplitude and observables}
\label{subsec:Transition ampl}

The second-order transition amplitude for Rayleigh (and Raman) scattering is given by \cite{L.Safari:12,Filippo:11,Akhiezer:65}
\begin{eqnarray}
\label{Mamplitude}
            \mathcal{M}_{if}(\uvector\epsilon_1,\uvector\epsilon_2)& =&
              \sumint{\nu} \frac{\bra{f}\mathcal{R}^{\dagger}(\vec k_2,\uvector\epsilon_2)
                \ket{\nu}\bra{\nu}\mathcal{R}(\vec k_1,\uvector\epsilon_1)\ket{i}}{\omega_{\nu i}-\omega_1}  \nonumber\\
           & & \hspace{-1cm} +  \sumint{\nu}\frac{\bra{f}\mathcal{R}(\vec k_1,\uvector\epsilon_1)
                \ket{\nu}\bra{\nu}\mathcal{R}^{\dagger}(\vec k_2,\uvector\epsilon_2)\ket{i}}{\omega_{\nu i}+\omega_2} \, ,
\end{eqnarray}
where $\omega_{\nu i} = (E_{\nu}-E_i)/\hbar$ is the transition frequency between states $\ket{\nu}$ and $\ket i$.
Here, the transition operator $\mathcal{R}(\vec k_{1(2)},\uvector\epsilon_{1(2)})$ describes the relativistic interaction between the bound electron and the incident (scattered) photon. 
In the Coulomb gauge, the explicit expression of this transition operator is
\begin{eqnarray}
\label{R operator}
    \mathcal{R}(\vec k,\uvector\epsilon)=\bm{\alpha}\cdot\uvector{\epsilon} 
    \, e^{i\vec k\cdot \vec r} ~,
\end{eqnarray}
where $\bm{\alpha}$ is the vector of Dirac matrices.

The summation over the intermediate states showed in Eq.~(\ref{Mamplitude}) runs over the complete one-particle spectrum $\ket{\nu}$, including a summation over the discrete part of the spectrum as well as the integration over the positive and negative energy continua.
Performing this summation is perhaps the most difficult part in determining the transition amplitude and we shall postpone further discussion on it to Sec.~\ref{sec:Comput}.

The initial $\ket i$ and final $\ket f$ states of atomic hydrogen have well-defined angular momentum $j$, angular momentum projection $m_j$ and parity $(-1)^l$, where $l$ is the orbital angular momentum of the larger component of the Dirac spinor.
In the following, we denote them respectively as $\ket{\beta_i,j_i,m_{j_i}}$ and 
$\ket{\beta_f,j_f,m_{j_f}}$, where $\beta$ is a collective label used to denote all the additional quantum numbers needed to specify the atomic states but for $j$ and $m_j$. For hydrogenic ions, specifically, $\beta$ refers to the principal quantum number $n$ and the parity quantum number $l$. Owing to the conservation of energy, moreover, the energies 
$E_{\gamma_{1,2}}$ and $E_{f,i}$ are simply related by the equation
\begin{equation}
\label{conservation of energy}
   E_f-E_i = E_{\gamma_1}-E_{\gamma_2} ~.
\end{equation}
Since Rayleigh scattering is an elastic process, the initial and final states coincide, $\ket{i}\,=\,\ket{f}$, and, thus, Eq.~(\ref{conservation of energy}) simplifies to $ E_{\gamma_1}=E_{\gamma_2}\equiv E_{\gamma}$.
It is furthermore assumed that both initial ($m_{j_i}$) and final ($m_{j_f}$) polarizations of atomic states remain unobserved, as typical for most experiments.

In this work, we shall separately consider the two experimental scenarios corresponding to circularly and linearly polarized light. 
In the {\it circular polarization scenario}, the incident light is circularly polarized or unpolarized and the polarization of the scattered light is measured in the circular base. In the {\it linear polarization scenario}, the incident light is linearly polarized or unpolarized and the polarization of the scattered light is measured in the linear base. We conduct our analysis by investigating the PDAD (i.e., the polarization-dependent angular distribution) and the degrees of circular and linear polarization of the scattered light.

For the circular polarization scenario, the PDAD can be
written in terms of the scattering amplitude \eqref{Mamplitude} as \cite{L.Safari:12}
\begin{equation}
\begin{array}{lcl}
\label{eq:dsig/domega_cc}
\frac{\dd \sigma^c}{\dd \Omega}(\lambda_1, \lambda_2, E_\gamma, \theta)&=&
\frac{\alpha^2c^2}{(2j_i+1)}\sum_{\substack{m_{j_i}m_{j_f}}}         \Big|\mathcal{M}^c(\lambda_1,\lambda_2)\Big|^2 ~,
\end{array}
\end{equation}
if the polarization of incident light is known.
In the above equation, $\alpha$ is the electromagnetic coupling constant, $c$ is the speed of light in vacuum and for simplicity we defined
\begin{equation}
\mathcal{M}^{c}(\lambda_1,\lambda_2)\equiv\mathcal{M}_{if}(\uvector\epsilon_{\lambda_1}^c,\uvector\epsilon_{\lambda_2}^c)~.
\end{equation}
If the incident light is unpolarized, the PDAD for the circular polarization scenario is obtained by taking \eqref{eq:dsig/domega_cc} and averaging over the (two independent) circular polarizations of the incident light \cite{F.Fratini-Hyrapetyan:11}:
\begin{equation}
\label{eq:dbarsig/domega_cc}
\frac{\dd\bar\sigma^c}{\dd\Omega}(\lambda_2, E_\gamma, \theta)=\frac{1}{2}\sum_{\lambda_1}\frac{\dd\sigma^c}{\dd\Omega}(\lambda_1,\lambda_2,E_\gamma, \theta)~.
\end{equation}
The PDADs for the linear polarization scenario, $\frac{\dd\sigma^l}{\dd\Omega}(\chi_1,\chi_2,E_\gamma, \theta)$ and $\frac{\dd\bar\sigma^l}{\dd\Omega}(\chi_2,E_\gamma, \theta)$, are simply obtained from \eqref{eq:dsig/domega_cc} and \eqref{eq:dbarsig/domega_cc} respectively, with the replacements $c\to l$ and $\lambda_{1,2}\to \chi_{1,2}$.

The degrees of circular ($P_C$) and linear ($P_L$) polarization of the scattered light, if incoming light is polarized, are defined as \cite{V. Balashov:00, S. Tashenov:06}
\begin{equation}
\label{eq:DegreeCircLin}
\begin{split}
P_C&=P_3~,\\
P_L&=\sqrt{(P_1)^2+(P_2)^2}~,
\end{split}
\end{equation}
where 
\begin{equation}
\begin{split}
P_1&=     \frac{\dd\sigma^l(\chi_1,0^\circ,E_\gamma, \theta)
          -\dd\sigma^l(\chi_1,90^\circ,E_\gamma, \theta)}
          {\dd\sigma^l(\chi_1,0^\circ,E_\gamma, \theta)
          +\dd\sigma^l(\chi_1,90^\circ,E_\gamma, \theta)}~,\\[0.4cm]
P_2&=
           \frac{\dd\sigma^l(\chi_1,45^\circ,E_\gamma, \theta)
          -\dd\sigma^l(\chi_1,135^\circ,E_\gamma, \theta)}
          {\dd\sigma^l(\chi_1,45^\circ,E_\gamma, \theta)
          +\dd\sigma^l(\chi_1,135^\circ,E_\gamma, \theta)}~,\\[0.4cm]
P_3&=
           \frac{\dd\sigma^c(\lambda_1,+1,E_\gamma, \theta)
          -\dd\sigma^c(\lambda_1,-1,E_\gamma, \theta)}
          {\dd\sigma^c(\lambda_1,+1,E_\gamma, \theta)
          +\dd\sigma^c(\lambda_1,-1,E_\gamma, \theta)} ~.\\
\end{split}
\end{equation}
$P_{1,2,3}$ are called the first, second and third Stokes parameter, respectively. We have denoted, above and in the following, $\dd\sigma/\dd\Omega$ by $\dd\sigma$ for brevity. If the incoming light is unpolarized, the degrees of circular and linear polarization of the scattered light will be denoted by $\bar P_C$ and $\bar P_L$, respectively. Their definitions are given by
\begin{equation}
\label{eq:DegreeCircLinAveraged}
\begin{split}
\bar P_C&=\bar P_3~,\\
\bar P_L&=\sqrt{(\bar P_1)^2+(\bar P_2)^2}~,
\end{split}
\end{equation}
where 
\begin{equation}
\label{Stokes}
\begin{split}
\bar P_1&=
           \frac{\dd\bar\sigma^l(0^\circ,E_\gamma, \theta)
          -\dd\bar\sigma^l(90^\circ,E_\gamma, \theta)}
          {\dd\bar\sigma^l(0^\circ,E_\gamma, \theta)
          +\dd\bar\sigma^l(90^\circ,E_\gamma, \theta)}~,\\[0.4cm]
\bar P_2&=
           \frac{\dd\bar\sigma^l(45^\circ,E_\gamma, \theta)
          -\dd\bar\sigma^l(135^\circ,E_\gamma, \theta)}
          {\dd\bar\sigma^l(45^\circ,E_\gamma, \theta)
          +\dd\bar\sigma^l(135^\circ,E_\gamma, \theta)}~,\\[0.4cm]
\bar P_3&=
           \frac{\dd\bar\sigma^c(+1,E_\gamma, \theta)
          -\dd\bar\sigma^c(-1,E_\gamma, \theta)}
          {\dd\bar\sigma^c(+1,E_\gamma, \theta)
          +\dd\bar\sigma^c(-1,E_\gamma, \theta)} ~.\\
\end{split}
\end{equation}

Prior to showing the results for the above defined PDADs and degrees of linear and circular polarization of the scattered light, in the following subsections we will further evaluate the scattering amplitude \eqref{Mamplitude} for both circular and linear polarization scenarios. We will show that the use of Wigner-Racah algebra technique allows for significant simplifications.

%
%
%
%
%
\subsection{Evaluation of the transition amplitude for circular polarization scenario}
\label{sec:circ}

We shall here evaluate the amplitude \eqref{Mamplitude} for the case in which the incident light is circularly polarized or unpolarized and the polarization of the scattered light is measured in the circular base. To this end, we expand the vector plane wave $\uvector\epsilon_\lambda^c e^{i\vec k\cdot \vec r}$ in terms of spherical tensors with well defined angular momentum properties \cite{M.E. Rose:57}:
\begin{eqnarray}
\label{multi-pole decomposition final}
              {\uvector\epsilon}_{\lambda}^c e^{i\vec k\cdot \vec r} &=&
              \sqrt{2\pi}\sum^{+\infty}_{L=1}\sum^{L}_{M=-L}\sum_{p=0,1}  
                  i^L[L]^{1/2} (i\lambda)^p\,\bm{a}^p_{LM}(k,\vec r) \nonumber\\
              & & \times\, D^L_{M\lambda}(\varphi_k,\theta_k,0)    ~,
\end{eqnarray}
where $[L_1,L_2,...,L_n]=(2L_1+1)(2L_2+1)...(2L_n+1)$, and the spherical tensor $\bm{a}^p_{LM}(k,\vec r)$ refers to the magnetic ($p=0$) and electric ($p=1$) multipoles.
Each term $\bm{a}^p_{LM}(k,\vec r)$ has angular momentum $L$, angular momentum projection $M$ and parity $(-1)^{L+1+p}$.

As seen from Eq.~(\ref{multi-pole decomposition final}), the angular dependence of the vector plane wave results from the Wigner (rotation) matrices. The Wigner matrices transform each multipole field, with original quantization axis along the photon propagation direction, into the field with quantization axis along the $z|| \vec k_1 $ direction (as showed in Sec. \ref{sec:Geometry}). This definition of the quantization axis enables us to describe the second photon direction by means of the single polar angle $\theta$. Thus the Wigner rotation matrices simplify as
$D^{L_2}_{M_2\lambda_2}(\varphi_{\vec k_2},\theta_{\vec k_2},0)= d^{L_2}_{M_2\lambda_2}(\theta)$ and $D^{L_1}_{M_1\lambda_1}(\varphi_{\vec k_1},\theta_{\vec k_1},0)= \delta_{M_1\lambda_1}$.

Combining Eqs.~\eqref{Mamplitude}, \eqref{R operator} and \eqref{multi-pole decomposition final}, and making use of the Wigner-Eckart theorem \cite{Sakurai:94}, the transition amplitude can be written as
\begin{widetext}
\begin{equation}
\label{eq:M}
\begin{array}{lcl}
       \ds\mathcal{M}^c(\lambda_1,\lambda_2)&=&
        2\pi\sum_{\substack{L_1L_2\\M_2}}\sum_{p_1p_2}(+i)^{L_1-L_2+p_1+p_2}
        [L_1,L_2]^{1/2}(\lambda_1)^{p_1}(\lambda_2)^{p_2}
        d^{L_2}_{M_2-\lambda_2}(\theta)
\\[0.8cm]
   &&   \times\;\sum_{j_\nu}
        (-1)^{-j_{\nu}}\frac{1}{(2j_{\nu}+1)^{1/2}}
        \Big(\Theta^{j_{\nu}}(1,2)S^{j_{\nu}}(1,2)
        +\Theta^{j_{\nu}}(2,1)S^{j_{\nu}}(2,1)
        \Big)~,
\end{array}
\end{equation}
where the reduced (second-order) matrix element is given by
\begin{equation}
\label{eq:S_J}
 S^{j_\nu}(1,2)= 
\sum_{\beta_{\nu}}\frac{\redbra{\beta_i,j_i}\bm\alpha\cdot\bm a_{L_1}^{p_1}(k_1,\vec r) \redket{\beta_{\nu},j_{\nu}}
\redbra{\beta_{\nu},j_{\nu}}\bm\alpha\cdot\bm a_{L_2}^{p_2}(k_2,\vec r)\redket{\beta_{i},j_{i}}}
{\omega_{\nu i}+\omega_2}~,
\end{equation}
\end{widetext}
and $S^{j_\nu}(2,1)$ is obtained from \eqref{eq:S_J} by i) interchanging the label 1 with 2 and ii) replacing the positive sign in the denominator with a negative sign. This latter replacement is given by the fact that, while the second photon is emitted, the first photon is absorbed by the atom.
Following the notation used in Refs. \cite{S.P. Goldman:81, L.Safari:12}, in Eq. \eqref{eq:M} we have furthermore defined  
\begin{eqnarray}
\label{eq:ThetaS}
\Theta^{j_{\nu}}(1,2)&=&\sum_{m_{j_\nu}}(-1)^{m_{j_f}+m_{j_{\nu}}}(2j_{\nu}+1)^{1/2} \\\nonumber
     &\times& \left(\begin{array}{ccc}
     j_f&L_1&j_{\nu}\\
     -m_{j_f}&\lambda_1&m_{j_{\nu}}
    \end{array}
\right)\left(
\begin{array}{ccc}
j_{\nu}&L_2&j_{i}\\
-m_{j_{\nu}}&M_2&m_{j_{i}}
\end{array}
\right)~,
\end{eqnarray}
where $\Theta^{j_{\nu}}(2,1)$ is obtained from Eq.~\eqref{eq:ThetaS} by replacing $L_1\leftrightarrow L_2$ and $\lambda_1\leftrightarrow M_2$.

Equation (\ref{eq:M}) can be used to evaluate Rayleigh scattering by hydrogenic as well as many-electron systems. For the latter case, we must just replace hydrogenic states with states of many-electron systems. In this article, results are only given for Rayleigh scattering by atomic hydrogen.

%
%
%
%
%
\subsection{Evaluation of the transition amplitude for linear polarization scenario}
\label{sec:lin}

We shall here evaluate the amplitude \eqref{Mamplitude} for the case in which the incident light is linearly polarized or unpolarized and the polarization of the scattered light is measured in the linear base. To this end,
we make a decomposition of the vector plane wave $\uvector\epsilon_\chi^l e^{i\vec k\cdot \vec r}$ by using the relation \cite{M.E. Rose:57}
\begin{equation}
\uvector\epsilon_\chi^l= -\frac{1}{\sqrt{2}}\sum_{\lambda=\pm1} 
e^{-i\lambda\chi}\uvector\epsilon_{\lambda}^c 
\label{eq:linear base}
\end{equation}
together with Eq.~\eqref{multi-pole decomposition final}. By combining Eqs. \eqref{Mamplitude},  \eqref{R operator} and \eqref{eq:linear base}, we get a simple equation which relates the amplitudes for circular and linear polarization scenarios:
\begin{eqnarray}
\label{eq:ll ampl}
      \mathcal{M}^l(\chi_1,\chi_2)
      =\frac{1}{2}\sum_{\lambda_1\lambda_2}e^{-i\lambda_1 \chi_1}e^{i\lambda_2 \chi_2}
      \mathcal{M}^c(\lambda_1,\lambda_2)~,
\end{eqnarray}
where for simplicity we defined
\begin{equation}
\mathcal{M}^l(\chi_1,\chi_2)\equiv\mathcal{M}_{if}(\uvector\epsilon_{\chi_1}^l,\uvector\epsilon_{\chi_2}^l)~.
\end{equation}
The amplitude for the linear polarization scenario can be thus easily evaluated by using Eqs. \eqref{eq:ll ampl} and \eqref{eq:M}.
%
%
%
%
%
%
\section{Computation}
\label{sec:Comput}

In this section, we shall discuss how the reduced matrix element (\ref{eq:S_J}) is calculated in the present work.

During the last decade, various methods have been investigated for calculating the reduced matrix element (\ref{eq:S_J}) as well as the whole transition amplitude (\ref{Mamplitude}) 
(see Refs. \cite{Nganso:07, A. Costescu:2011, A. Costescu:94, F.Fratini-S.Trotsenko:11, A. Surzhykov:10}). In practice, the summation over the complete spectrum contained in Eq. \eqref{eq:S_J} cannot be performed explicitly. Several approaches and approximation techniques have thus been proposed in the literature to perform this summation, such as the Coulomb-Green function approach, which has been widely used for studying different decay and scattering processes of atoms and ions \cite{skf2005, M. Alfred:98}.

An alternative approach is the \textit{finite basis set} method \cite{saj1996, jbs1998,D. D. Bhatta:06, ZatBar2008, FroZat2009, A. Surzhykov:11, sfi1998}.
The finite basis set method is based on the supposition that the ion (or atom) is enclosed in a finite cavity. Such a restriction leads to a ``discretized'' continuum part of the atomic or ionic spectrum, and hence to a representation of the Dirac wavefunctions in terms of pseudo basis set functions \cite{jbs1998}. 
The radius of the cavity $R$ is of course taken large enough to ensure a good approximation for the wave functions.

In the present work, we calculate the reduced matrix element (\ref{eq:S_J}) by adopting the finite basis set method, using B splines and B polynomials as finite basis sets. The B splines are one of the most commonly used family of piecewise polynomials, since they are well adapted to numerical tasks \cite{jbs1998}. The B polynomials, or the Bernestein polynomials \cite{D. D. Bhatta:06}, are a good alternative to the B splines since they allow for analytical finite basis-set calculations. These are polynomial functions of $n\textrm{th}$ degree that are used to obtain the solution of linear and nonlinear differential equations \cite{D. D. Bhatta:06}. The details of these basis sets, as well as a comparison between them, can be found in Ref. \cite{asfis2011}. Thus, we restrict ourselves to describe the characteristic parameters used in this work.

The parameters of the B splines basis set are the radius of the cavity ($R_{\rm{bs}}$), the number of B splines ($n_{\rm{bs}}$) and their degree ($k$). As for the B polynomials, the parameters are the radius of the cavity ($R_{\rm{bp}}$) and the number of B polynomials ($n_{\rm{bp}}$) (the degree of the B polynomials is $n_{\rm{bp}}-1$).
The parameters used in both basis sets were optimized in order to obtain stability and agreement of six digits between the results of both basis sets. The optimal parameters are:  $R_{\rm{bs}}=60$~a.u., $n_{\rm{bs}}=60$, $k=9$, $R_{\rm{bp}}=50$~a.u. and $n_{\rm{bp}}=40$. Such set of parameters was already obtained for the case of two photon emission \cite{aspsi2009,asfis2011}, and for the angular distribution in Rayleigh scattering by atomic hydrogen \cite{L.Safari:12}.
%
%
%
%
%
%
%
\section{Results and discussion}
\label{sec:ResultsDiscussion}
In this section, we shall present the results for the PDADs and the degrees of circular and linear polarization of the scattered light which we defined in Sec. \ref{subsec:Transition ampl} (see Eqs. \eqref{eq:dsig/domega_cc}, \eqref{eq:dbarsig/domega_cc}, \eqref{eq:DegreeCircLin} and \eqref{eq:DegreeCircLinAveraged}). Such results have been obtained by using the relations presented in Secs. \ref{sec:circ} and \ref{sec:lin}, and by implementing the computation technique presented in Sec. \ref{sec:Comput}.
%
%
%
%
%
%
%
\subsection{Circular polarization scenario}
\label{circ,circ}

\begin{figure}[b]
\includegraphics[scale=0.73]{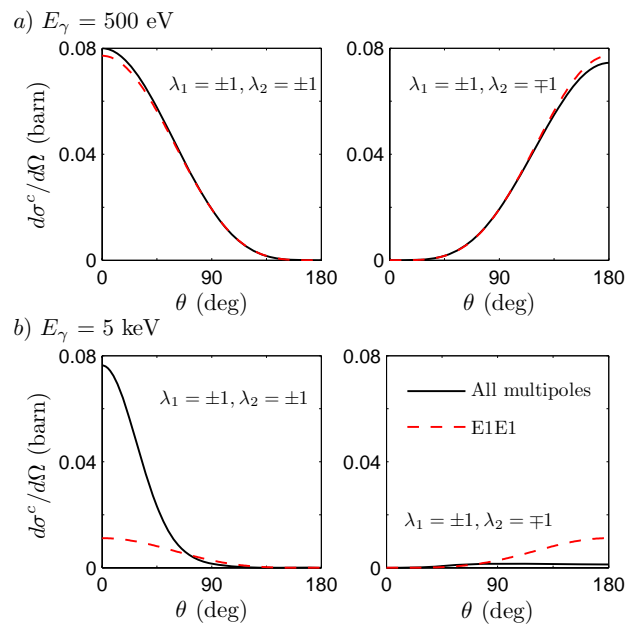}
\caption{(Color online) 
Angular distribution of the scattered light when the incident light is circularly polarized and the scattered light is measured in the circular base ($\dd\sigma^{c}/\dd\Omega$).
Results are calculated with the account of all photon multipoles (solid-black line) and within the electric dipole approximation (dashed-red line), for two selected photon energies: $a)$ $E_{\gamma}$= 500 eV; $b)$ $E_{\gamma}$= 5 keV.
}
\label{fig:2}
\end{figure}

Fig.~\ref{fig:2} displays the angular distribution function (\ref{eq:dsig/domega_cc}), $\dd\sigma^{c}/\dd\Omega$, for the four polarization configurations $\lambda_1=\pm1$, $\lambda_2=\pm1$ and $\lambda_1=\pm1$, $\lambda_2=\mp1$. The analyzed photon energies are 500 eV ($a$) and 5 keV ($b$). The shape of the angular distribution for $E_\gamma= 500$ eV  can be easily understood owing to the conservation of the angular momentum: Since the leading electric-dipole (E1E1) term is independent of spin interaction operators within the nonrelativistic limit, the photon helicity must be conserved (flipped) for forward (backward) scattering. 
A similar discussion also applies to the higher energy case $E_\gamma$ = 5 keV. However, as discussed elsewhere \cite{L.Safari:12},  multipoles beyond the E1E1 approximation determine here a strong suppression of backwards scattering, thereby suppressing helicity-flip scattering events.
\begin{figure}[t]
\includegraphics[scale=0.52]{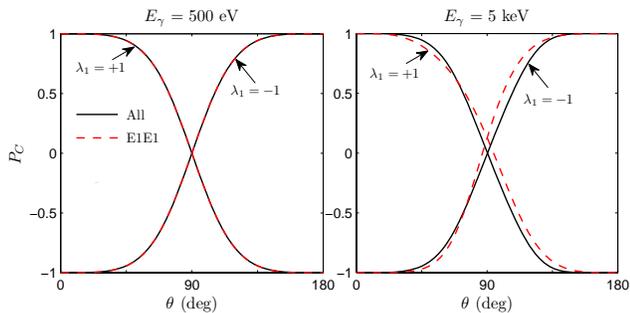}
\caption{(Color online) 
Degree of circular polarization of the scattered light when incident light is circularly polarized ($P_C$), as a function of the scattering angle $\theta$. The curves corresponding to positive ($\lambda_1=+1$) and negative ($\lambda_1=-1$) helicity of the incident photon are displayed.
Results are calculated with the account of all photon multipoles (solid-black line) and within the electric dipole approximation (dashed-red line).}
\label{fig:3}
\end{figure}

Fig.~\ref{fig:3} displays $P_C$ (i.e., the degree of circular polarization of the scattered light), as a function of the scattering angle $\theta$, for the two energy values $E_\gamma$ = 500 eV and 5 keV. When compared with Fig. 2, it is evident that the degree of circular polarization is less sensitive to corrections from higher multipoles than the angular distribution. At $E_\gamma$ = 5 keV, the (weak) effect that higher multipoles have on the degree of circular polarization of the scattered light, $P_C$, is to slightly increase it within the interval $0\le\theta\lesssim60^\circ$, where 90\% of the scattering events occur \cite{L.Safari:12}.

The angular distribution function $\dd\bar\sigma^{c}/\dd\Omega$ can be obtained from Fig.~\ref{fig:2}, by taking the arithmetic mean of the curves referring to positive ($\lambda_1=+1$) and negative ($\lambda_1=-1$) helicity of the first photon, for fixed $\lambda_2$.
Following this procedure, the angular distribution takes the familiar shape $\sim1+\cos^2\theta$ at low energies, independently of the helicity of the scattered photon.

The function $\bar P_C$ (i.e., the degree of circular polarization of the scattered light for unpolarized incident light) vanishes for all angles and energies. This is easily seen by applying the definition \eqref{eq:DegreeCircLinAveraged} to the graphs showed in Fig.~\ref{fig:2}.
Such a result is somehow expected since, in this case, both hydrogen atom and incident light are unpolarized and therefore there cannot be any preferred direction for the circular polarization of the scattered light. If there were any, then violation of parity would occur.

%
%
%
%
\subsection{Linear polarization scenario}

\begin{figure}[t!]
\includegraphics[scale=0.73]{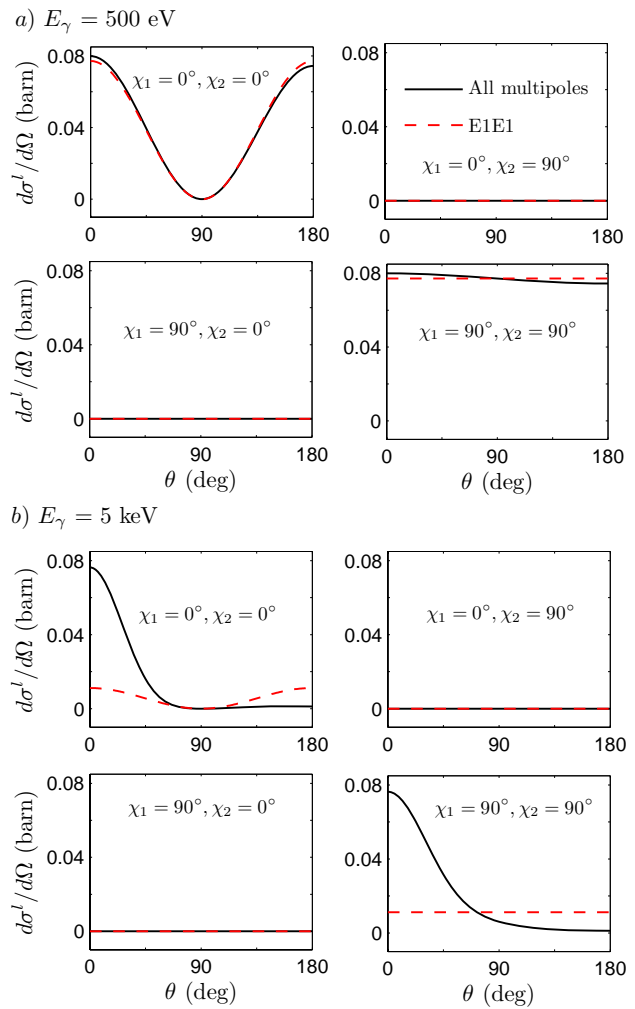}
\caption{(Color online) 
Angular distribution of the scattered light when the incident light is linearly polarized and the scattered light is measured in the linear base ($\dd\sigma^{l}/\dd\Omega$). Results are calculated with the account of all photon multipoles (solid-black line) and within the electric dipole approximation (dashed-red line), for two selected photon energies: $a)$ $E_{\gamma}$= 500 $eV$; $b)$  $E_{\gamma}$= 5 $keV$.
}
\label{fig:4}
\end{figure}

Fig.~\ref{fig:4} displays the angular distribution function $\dd\sigma^{l}/\dd\Omega$ for the four polarization configurations $\chi_1=0^\circ (90^\circ)$, $\chi_2=0^\circ (90^\circ)$ and $\chi_1=0^\circ (90^\circ)$, $\chi_2=90^\circ (0^\circ)$. The analyzed photon energies are 500 eV ($a$) and 5 keV ($b$). Similarly to the circular polarization scenario, deviations from the electric-dipole approximation are more pronounced for higher photon energies.
At low energies, where the electric-dipole approximation holds, we recover the well-known behavior
$\dd\sigma^{l}(\chi_1,\chi_2)\propto \big|\uvector\epsilon_{\chi_1}^{l}\cdot\uvector\epsilon_{\chi_2}^{l}\big|^2 $, which 
characterizes the polarization-dependent angular distribution in the non-relativistic (Thomson) limit. More specifically,
at low energies we obtain 
\begin{equation}
\begin{array}{c}
\dd\sigma^{l}(0^\circ,0^\circ)\propto\cos^2\theta ~,\quad \dd\sigma^{l}(90^\circ, 90^\circ)\propto \textrm{(const)} ~, \\[0.4cm]
\dd\sigma^{l}(0^\circ,90^\circ)=\dd\sigma^{l}(90^\circ, 0^\circ)=0 .
\end{array}
\end{equation}
These non-relativistic polarization correlations are the same as for two-photon decay in hydrogen atom \cite{F.Fratini:11_2}. This result is not unexpected in view of the similarity of the amplitude for the two processes.

As seen from Fig.~\ref{fig:4}, the angular distribution for Rayleigh scattering vanishes at normal angle ($\theta\approx90^\circ$) if  incident light is linearly polarized along $\chi_1=0^\circ$ direction. 
Moreover, the same figure shows that spin-flip scattering events (i.e., events for which $\chi_1=0^\circ$, $\chi_2=90^\circ$ or $\chi_1=90^\circ$, $\chi_2=0^\circ$) are strongly suppressed, at any angle $\theta$ and any energy $E_\gamma$. The linear polarization of the photon, which coincides with its spin, is therefore conserved, at any photon energy.

As can be easily verified from Eq. \eqref{eq:DegreeCircLin}, suppression of spin-flip transitions implies also that the degree of linear polarization, $P_L$, turns out to be simply $\approx1$, for incident polarization along $\chi_1=0^\circ$, $90^\circ$ direction, and therefore is not  displayed. 

The results for the function $\dd\bar\sigma^{l}/\dd\Omega$ can be obtained from Fig. \ref{fig:4}, by taking the arithmetic mean of the curves referring to $\chi_1=0^\circ$ and $\chi_1=90^\circ$, for fixed $\chi_2$.

\begin{figure}[t]
\includegraphics[scale=0.7]{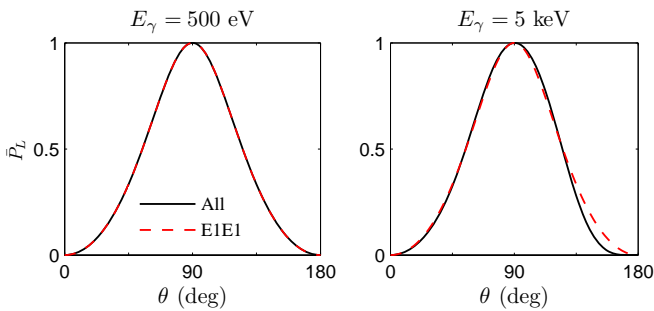}
\caption{(Color online) Degree of linear polarization of the scattered light when incident light is unpolarized ($\bar P_L$), as a function of the scattering angle $\theta$. Results are calculated with the account of all photon multipoles (solid-black line) and within the electric dipole approximation (dashed-red line).}
\label{fig:5}
\end{figure}

Fig.~\ref{fig:5} displays $\bar P_L$ (i.e, the degree of linear polarization of the scattered light for unpolarized incident light) as a function of the scattering angle $\theta$, for the two photon energies 500 eV and 5 keV. We read from the figure that light scattered at normal angle is fully linearly polarized, while light scattered forwards or backwards is unpolarized, at any photon energy. This polarization feature is also known to characterize the scattering of light by small molecules, when the photon wavelength is small compared with the molecular radius. It is due to this property that the sunlight scattered by surfaces at normal angles is always linearly polarized along the axis which is orthogonal to the scattering plane \cite{C.F.Bohren:85}.
%
%

%
%
%
%
%
\section{Summary and perspectives}
\label{sec:SumConcl}

We studied polarization properties of light in Rayleigh scattering by unpolarized atomic hydrogen, based on second-order perturbation theory and Dirac relativistic equation. 
We introduced two experimental scenarios in which the light is circularly and linearly polarized, respectively. For each of these scenarios, we analyzed the polarization-dependent angular distribution and the degrees of circular and linear polarization of the scattered light. To this end, we first decomposed the Rayleigh transition amplitude in terms of spherical tensors (angular part) and reduced amplitudes (radial part). We then calculated these latter for scattering by hydrogen atom by means of the finite basis-set method based on the relativistic Dirac equation. The polarization dependent angular distribution and the degrees of circular and linear polarization of the scattered light were plotted for the photon energy range 0.5 to 10 keV.

We found that, for circularly polarized incident light, helicity-flip scattering events are allowed at low energies ($E_\gamma\lesssim$ 500 eV) and are suppressed at high energies ($E_\gamma\gtrsim$ 5 keV), due to the suppression of backward scattering. Thus, the helicity of the incident photon is not conserved at low photon energies while is conserved at high photon energies.

For linearly polarized incident light, it was showed that spin-flip scattering events (i.e., events for which the linear polarization angle of incident and scattered light differ by a normal angle) are strongly suppressed, at any scattering angle and photon energy. Thus, the linear polarization of the incident photon is approximately conserved and conveyed to the scattered photon, in the whole investigated photon energy range.

Finally, light scattered at normal angles was found to be fully linearly polarized, at any photon energy value.

\medskip

With advances in polarization sensitive detectors in the x-ray and $\gamma$-ray region,  theoretical studies on polarization of photons emitted or scattered by atomic systems have lately become important \cite{Bondarev, S. Tashenov:06, S. Tashenov:09}. Further studies to investigate polarization properties of light scattered by solid targets are underway. 
\section{acknowledgment}
L. S. and F. F. acknowledge financial support by the Research Council for Natural Sciences and Engineering of the Academy of Finland. 
P. A. and S. T. acknowledges the support of German Research Foundation (DFG) within the Emmy Noether program under Contract No. TA 740 1-1. 
S. F. acknowledges support by the FiDiPro program of the Finnish Academy.
J. P. S. and P. A. acknowledge the support by FCT -- Funda\c{c}\~ao para a Ci\^encia e a Tecnologia (Portugal), through the Projects No. PEstOE/FIS/UI0303/2011 and PTDC/FIS/117606/2010, financed by the European Community Fund FEDER through the COMPETE -- Competitiveness Factors Operational Programme. 
F.F. acknowledges financial support by Funda\c{c}\~ao de Amparo \`a Pesquisa do estado de Minas Gerais (FAPEMIG).

%
%
%
%
%
%
\end{document}